Гусманов К.М. [1], Ханда К.Р. [1], Салихов Д. Б. [1], Маццара М. [1], Мавридис Н. [1]

[1] Университет Иннополис, г.Иннополис, Россия

# JOLIE GOOD BUILDINGS: ИНТЕРНЕТ ВЕЩЕЙ ДЛЯ ИНФРАСТРУКТУРЫ УМНЫХ ЗДАНИЙ С ИСПОЛЬЗОВАНИЕМ ПАРАЛЛЕЛЬНЫХ ПРИЛОЖЕНИЙ И РАСПРЕДЕЛЕННЫХ МИКРОСЕРВИСОВ

**АННОТАЦИЯ**

Большое число зданий, бытового или специального назначения, становятся "умнее" , в связи с огромными преимуществами с точки зрения экономии энергии, безопасности, гибкости и комфорта. Однако, на данный момент нет явно доминирующего общепринятого фреймворка для работы с аппаратным и программным обеспечением различных устройств. В данной работе мы представим прототип платформы для поддержки нескольких параллельных приложений для «умных» зданий, которая использует сети датчиков и распределенную архитектуру микросервисов, разработанную на языке программирования Jolie. Также мы расскажем про архитектуру и преимущества нашей системы и продемонстрируем прототип, работающий с несколькими устройствами, содержащий в себе пользовательский интерфейс и установленный на территории нашего кампуса. Полученные результаты наглядно демонстрируют перспективность нашего подхода и открывают новые преспективы для дальнейшей работы.

**КЛЮЧЕВЫЕ СЛОВА**

Умные здания; Интернет вещей; Управление энергией; Микросервисы; Jolie; Распределенные архитектуры; Параллельные приложения; Облачные вычисления

**Gusmanov K.M. [1], Khanda K.R. [1], Salikhov D. B. [1], Mazzara M. [1], Mavridis N. [1]**

[1] Innopolis University, Innopolis, Russia

# JOLIE GOOD BUILDINGS: INTERNET OF THINGS FOR SMART BUILDING INFRASTRUCTURE SUPPORTING CONCURRENT APPS UTILIZING DISTRIBUTED MICROSERVICES

**ABSTRACT**

A large percentage of buildings, domestic or special-purpose, is expected to become increasingly "smarter" in the future, due to the immense benefits in terms of energy saving, safety, flexibility, and comfort, that relevant new technologies offer. However, concerning the hardware, software, or platform levels, no clearly dominant standard frameworks currently exist. Here, we will present a prototype platform for supporting multiple concurrent applications for smart buildings, which is utilizing an advanced sensor network as well as a distributed micro services architecture, centrally featuring the Jolie language. The architecture and benefits of our system are discussed, as well as a prototype containing a number of nodes and a user interface, deployed in a real-world academic building environment. Our results illustrate the promising nature of our approach, as well as open avenues for future work towards it wider and larger scale applicability.

## INTRODUCTION

The Internet of Things (IoT), as per the ITU Recommendation ITU-T Y.2060 [1], has been defined as: "a global infrastructure for the information society, enabling advanced services by interconnecting (physical and virtual) things based on existing and evolving interoperable information and communication technologies." One such important category of physical things are buildings; functioning either as human habitats, for the case of domestic buildings, or specialized towards many other goals, as is the case for storehouses, shops, industrial buildings, schools, and much more. There are multiple aspects of the operation of modern-day buildings that afford further automation and optimization: for example, it has been shown that the benefits of better energy management through automation can be very high [2]; also, the security of buildings, their human-friendliness and adaptation to preferences can be vastly improved, not to mention other aspects. Thus, smart buildings, as well as their wider container of smart cities, are not only significant research topics for today, but also promise to improve our lives and increase sustainability, starting in the near future.

Traditionally, most building automation systems were made for specific applications and offered little degree of openness and flexibility. However, with the fast maturation of a number of supporting technologies, the opportunity to change this status quo is rapidly growing. First, cheap sensing and perception technologies have become available for a wide range of measurables: covering not only physical properties of the building and its spaces, such as temperature, light, and humidity, but also providing information about the presence, number, identities, activities, and even emotional states of the people inside a building or in its surrounding spaces. Second, affordable and miniature microprocessor-based platforms have become widespread and are easily inter-connectable to sensors, which often have enough processing power to support perception and machine vision; with multiple network transports, even the necessary small battery power is readily available. Third, networking technology for such platforms has advanced significantly, and nowadays it is easy to implement building-wide networks, often with dynamic ad-hoc topologies, which also support on-the-fly introduction and replacement of new nodes, while providing secure communications. Fourth, special languages and middleware has been developed, in order to support straightforward development of distributed systems based around a number of paradigms, including microservices.

Given these developments, one cannot only envision, but can also start implementing and experimenting with, the usage of the IoT for providing generic smart building infrastructure. Such an infrastructure would go beyond the constraints of existing specialized systems, and could provide fluid support for diverse concurrent applications, which would effectively share the resources of the infrastructure, and which would operate through microservices provided by the nodes. Furthermore, infrastructure such as the "Jolie Good Buildings" that we present here, promises strong scalability, reliability, and upgradability as more powerful hardware becomes available. Yet another important advantage of our system, is its support for the direct utilization of the immense power of external services available on the wider internet, beyond its own hardware, as part of the distributed applications running on its nodes.

In this paper, we will start by providing background on relevant existing work. Then, we will present the overall architecture of our system and will explain the requirements and design choices made. We will then present results, discuss current and future steps, and finish with a forward-looking conclusion. We hope that this work will help create a future, in which not only resources are saved and the environment protected, but also human life becomes less stressful with enhanced efficiency and creativity.

**BACKGROUND**

The Internet of Things (IoT) promises to become an integral and important part of our future. At the moment, the number of connected devices is greater than the estimate of 22 billion [3], which is several times greater than the number of Earth's inhabitants. However, one of the problems that arises is the existence of many different types of incompatibilities among devices, prohibiting fluid interoperability, when in many cases what is needed for specific applications is already available in distributed form around the IoT. For example, consider the kinds of information one can get from a fitness tracker device: using the accelerometer and gyroscope, one can get information about whether the person wearing the device is now running, walking, eating or swimming in a pool. However, the most important issue is the ability to integrate this information with other information existing in other sources. For example, on the basis of data from a heart rate sensor, as combined with fitness tracking information and personal data, a smart application could start assessing the health condition of an individual, and if required, suggesting various treatments. Also, such information could also be very useful for professional doctors - a physician could utilize such extensive and detailed historical data effectively: understanding the temporal trajectory of a health issue, the frequency of observable problems etc.

The IoT can be envisioned to be useful in almost all spheres of human life. In our project we consider the theme of Smart Buildings and Human-Building Interaction. There have been many different studies on this topic, which utilize various kinds of sensors, software architectures and, sometimes also include robots. The creation of a single and simple platform that can handle a large variety of types of sensors and even robots is a very important step towards the development of Smart Buildings. It should be noted that Cloud Computing (CC) could also play a beneficial role in such systems. In the words of [4]: "In cloud computing, computation is viewed as a "utility". In a similar sense with modern power and water networks, cloud users do not need to own the means of production or distribution (i.e. power generators, water sources and distribution networks); they just connect to the cloud, and time-share reusable distant distributed computation, storage, and code resources that become available in a transparent fashion (not knowing the whereabouts or the specifics of them) and with high robustness.". Yet, traditional Cloud Computing also has a number of limitations. One of them is that "it is not effectively connected to the physical world in the way that a situated robotic agent would be" [7], and this is a prime motivation for the introduction of the concept of the Human-Machine cloud, which goes beyond platform-as-a-service, code-as-a-service, and storage-as-a-service, by including sensing, cognition, and actuation services, provided through both humans and machines.

A further stage of the development of the system of Smart Buildings is to design effective interaction between humans and smart buildings. A concise overview of some relevant research can be found in [5]. Furthermore, an initial example of a framework for human-machine cloud services can towards smart buildings, can be found in [6].

Microservices can be very useful towards the IoT. Microservice software architecture is a style for service-oriented development whose popularity is increasing now because of growing interest to parallel and distributed computation. Designed and built especially towards microservices, the Jolie programming language [8] was developed to work directly with a service-oriented paradigm, which distinguishes it from other popular languages like C#, Java, or Python. This means that the language contains features that are unique to this approach, i.e. example representation of building blocks. In object-oriented languages, there are usually classes or functions; in Jolie, the building blocks are a service in their own right.

**ARCHITECTURE**

In this stage, the main goal was to create a small and simple system, where the biggest part of the work was assigned to Jolie (Figure 1). All process was divided by steps and as the result the goal was achieved.

The first step was connecting and configuring sensors through BLE. The main data collected for temperature, humidity and luminance were made using CC2650 SensorTag because they have the necessary sensors, small size, run on batteries, and easy code to write. The work with sensors are very similar to the creation of the sketches for Arduino: C code is written with the necessary libraries for the SensorTag, sensors to be used, and ways of exchanging information with these devices. In our case, data is sent via BLE, as it is very easy to use and configure.

The door sensor from Aeon Labs, which works on the protocol Z-Wave, was used to create a monitoring system entrance/exit to the premises. To work with this device, HomeOS[6] was used, which is written in the programming language C# and has the necessary functionality to work with the devices of this company in the protocol Z-Wave.

The next step was to writing simple code to work with the sensors in Jolie. What are the advantages of using Jolie? The most interesting opportunity that we receive is code reusability. Because our system will support different types of sensors, the main logic of their connectivity and data extraction is similar for most of them. Also, product support becomes easier: this is reusability. The same service can be used for these sensors. Implied by the first advantage, the second advantage is reducing bugs. A third advantage that may be interesting for future collaborators is simplicity. Jolie divides all system logic into small parts: we have several services that are responsible for each sensor or each action, the naming of each block is intuitively understandable, so these language features increase code readability. In addition, one more significant advantage is working with Java code in Jolie. Thus, the part of the work with getting data from BLE devices was written in the Java programming language and divided into simple functions, e.g. connecting to devices and data retrieval, which is called from Jolie. Thus, the code is easy to read, clear, and ready for further work as a client for other sensors.

The last step is data collection. From SensorTags, we read data about room temperature, humidity and light. All of them were recorded in .csv format for later processing and graphing in Python. Data from Aeon Labs sensor is the number of opening/closing doors and monitoring incoming/outgoing people using the webcam, which recorded in .csv and .jpg format respectively. It is also worth noting that all codes worked in Raspberry Pi, which has all necessary libraries and connections.

Despite its apparent simplicity, there were several issues at work with devices. The first serious problem was the lack of libraries for Java to work with BLE devices. But we found a great and simple library for Intel Edison devices and used it. The next issue was working with Z-Wave devices in HomeOS, which periodically generated exceptions and did not connect to the device. However, all these problems were solved, which allowed us to come to the result, which will be described in the next section.

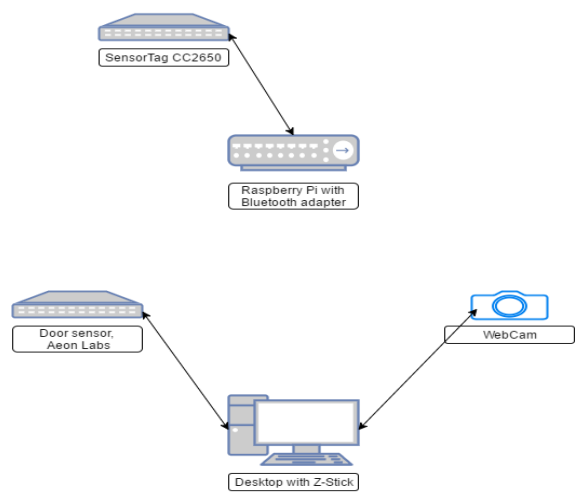

*Figure 1. A scheme of devices and their connections*

**RESULTS**

We have developed a platform to work with the SensorTags based on Jolie and Java programming languages.

Despite its small size, it obtained most of the data, with which we work. The ability to embed Java code allows the possibility to add necessary functions and give more functionality and usability to Jolie language. At this stage, our goal was to build the basic functionality of working with sensors, understanding how they work and how to interact with them and get ideas for the further development of the project. As a check on the performance, we have placed sensors in two rooms – in the student's room in the dormitory and in the laboratory, in which we spend most time of the day. It is worth noting that both rooms can be ventilated and fitted with lights, heaters, special motors and actuation mechanisms to control air flow and solar illumination etc.

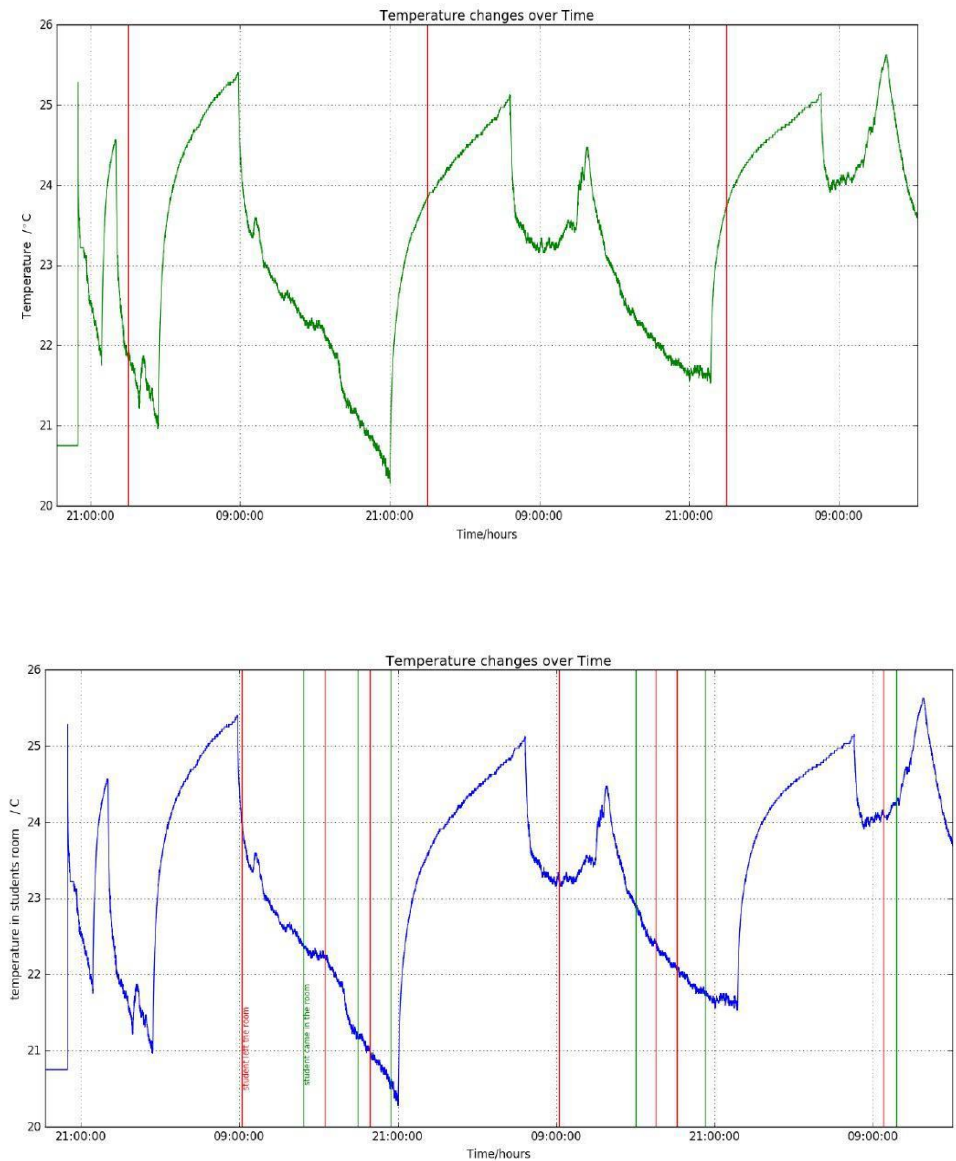

*Figure 3. A graph of the temperature in the room of a student with periods of absence in the room (red & green lines)*

In the room of the student, we set the Raspberry Pi with the connected Bluetooth adapter and the SensorTag to obtain data on the temperature. For tracking of location of the student in the room, we used data from a fitness-tracker MiBand2, using its mac address. Since the room is small, then we could accurately determine when the student arrives and when he leaves. Based on this dataset was constructed a graph of the temperature in the room (Figure 2, 3), which will later be used to make up the optimal temperature in the room of the student.

In the laboratory, we placed more devices and used multiple platforms to work with them. This time SensorTag sent data about room temperature, humidity and light. Also, we used a sensor of opening/closing doors and a web camera to record those entered/exited the room. As noted earlier, for obtaining data on temperature, humidity and light, their processing and saving to file, we used two programming language – Java and Jolie, which were running on the Raspberry Pi. Work with other sensors and a camera was assigned to HomeOS, which has the necessary functionality for working with the devices. Based on all the collected data, we constructed graphs of the temperature, humidity, and light conditions (Figure 4, 5, 6). Using a sensor on the door and the camera data, we obtained data how many people were in the room in each moment of time and assembled a small dataset of photo lab workers.

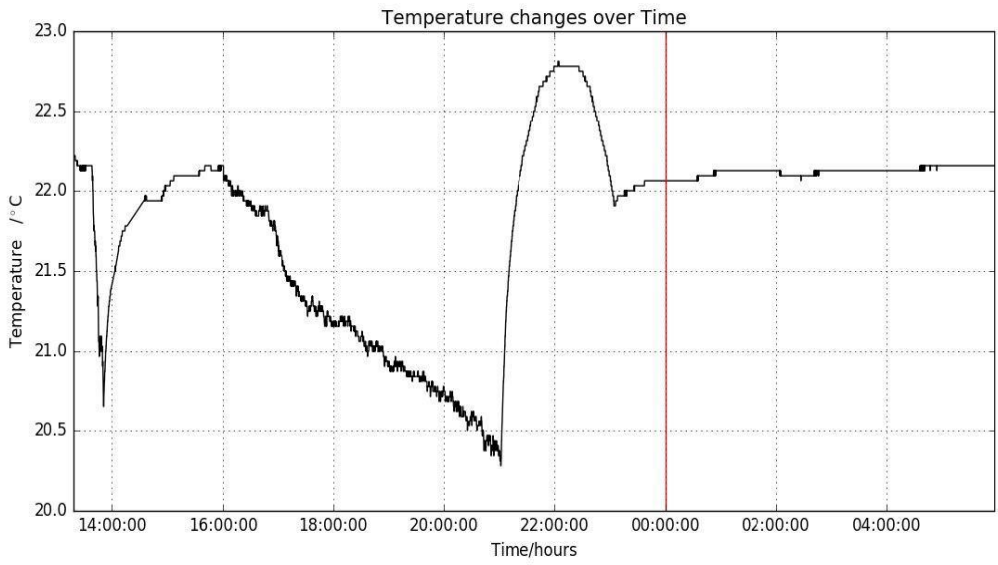

*Figure 4. A graph of the temperature in the lab*

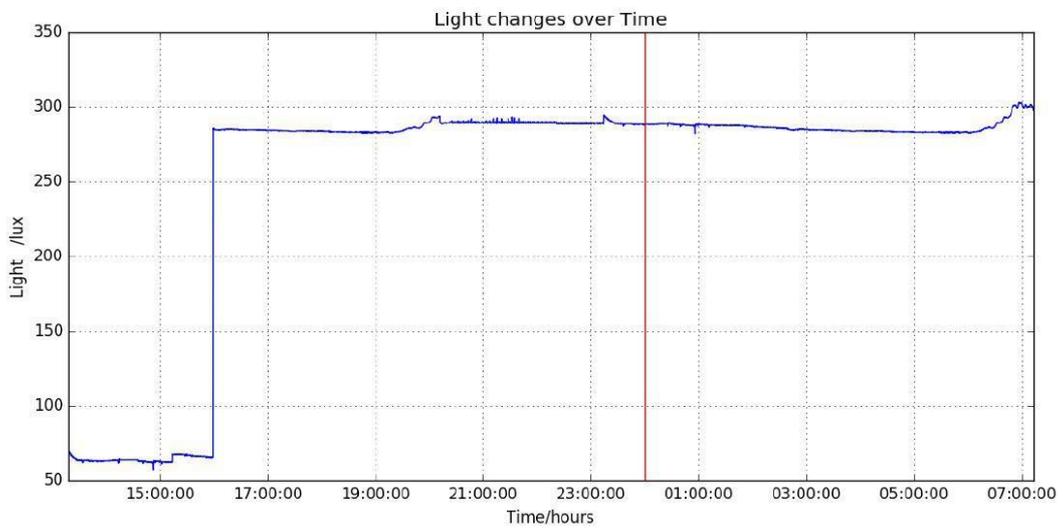

*Figure 5. A graph of the light in the lab*

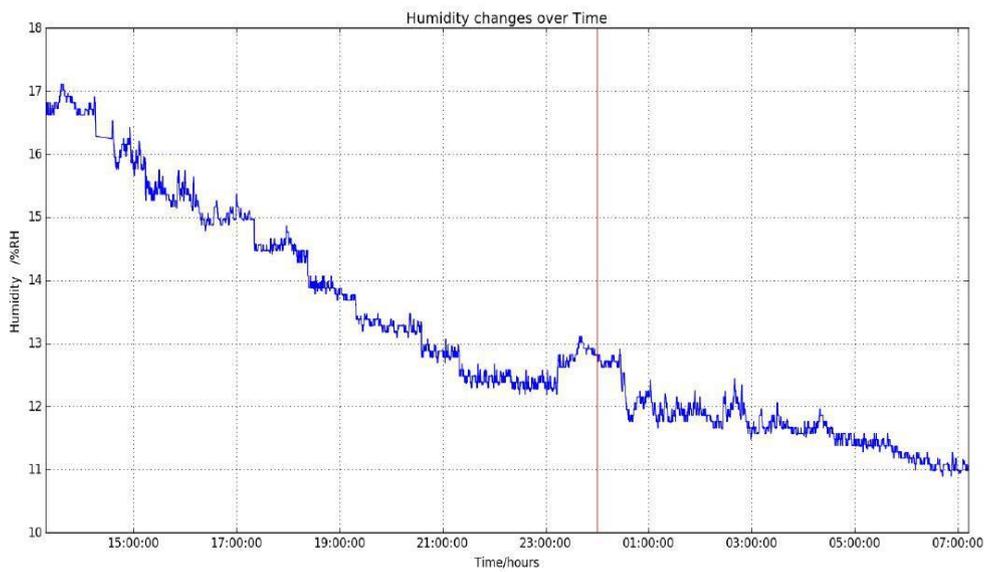

*Figure 6. A graph of the humidity in the lab*

This completes the main work of the first stage. Plans for further development of the project and data will be described in the next section.

**FUTURE STEPS**

Despite having solved many of the problems with obtaining data from the sensors and further processing, in order to get the targeted result there are many problems that need to be addressed. First, we must configure all SensorTags to work on the ZigBee protocol because in the future we will set up these sensors in a variety of classrooms at our university, and there is a need to build a mesh network for communication between devices and further processing in the language Jolie.

Second, we need to expand the functionality of the language Jolie to work with all our devices. In this case, we must write the modules for BLE, ZigBee and Z-Wave devices to continue to connect easily and share data between smart devices. In addition, we want to rewrite the module for working with the Aeon Labs sensors in Jolie language.

Thirdly, we need to place all sensors and the Raspberry Pi in the classroom of the University and in the room of the student and start collecting data. In the room of the student, they will continue collecting data on temperature including the data on humidity and lighting as well as installing a window-opening sensor that will allow us to get good dataset to determine the preferred mode of temperature, light and humidity in the room and drawing a small daily schedule of a student. In the classrooms, we will also install a small camera to track the number of people in the room and launch a small bot to Telegram messenger to gather feedback from students in terms of comfort temperature and humidity in the auditorium. In the lab, we will add a motion sensor and several relays to control lights and power sockets in the room that will be activated either automatically or manually through a central control system. One could even envision integration of our system with outdoor area monitoring (for example, car park areas [11]), vegetation monitoring and identification [12], or with recommendation systems in order to create user-tailored hospitality services in large buildings [13], and with information based on periodic aerial monitoring [14].

Based on the obtained data, which will take about a month to collect, we will build a prediction system for students and employees, which will be checked by a simple small bot in Telegram. Thus, we will improve our result in order to begin to create a small artificial intelligence that is responsible for the control of lighting, temperature, electricity and comfort level in the room.

All this will lead to the creation of a unified platform, unified control center, written in Jolie that will allow:
1. To develop a system of microservices and the language Jolie;
2. Easy to incorporate new devices operating according to one of three protocols;
3. To track the preferences of students and staff in the environment;
4. To control the flow of electricity to the premises and, if possible, to suggest ways to reduce costs;
5. To use predictive methods on the basis of long-term logged data;
6. To create examples of concurrent applications supporting Energy Management, Security & Hospitality;
7. Develop the idea of Human-Building Interaction and technologies on campus and the entire city;

**CONCLUSION**

As the Internet of Things becomes more widespread, and given the current availability of sensors and embedded processing units, the prospect for advanced smart buildings utilizing such technologies becomes increasingly attractive. Furthermore, the possibility of creating architectures utilizing distributed microservices and supporting various concurrent applications to be running on them (such as energy management, security, and hospitality applications), is opening new avenues for flexibility, scalability and reusability. Towards that goal, we created a platform based on a network of advanced SensorTags utilizing Jolie: a new, simple yet flexible language, which supports microservices and the Internet of Things. Through an initial implementation, we demonstrated the viability and power of our approach, and we are also contributing towards creating an expanding set of tools in Jolie that will be shared and reused with our wider community. We are currently working on enlarging our network and data collection, and on integrating advanced visualizations and human-building interfaces, as well as prediction abilities and user-centered customization. We hope that through the wider introduction of such systems, the buildings of the coming decade will be much more interesting places to live and work, that will not only be efficient, but will also help preserve our environment, and create sustainable prosperity for mankind, in harmony with nature.

Об авторах:

**Гусманов Камилл Марселевич**, студент факультета компьютерных наук Университета Иннополис, k.gusmanov@innopolis.ru

**Ханда Кевин Раджнишевич**, студент факультета компьютерных наук Университета Иннополис, k.khanda@innopolis.ru

**Салихов Дильшат Булатович**, студент факультета компьютерных наук Университета Иннополис, d.salikhov@innopolis.ru

**Маццара Мануэль**, руководитель лаборатории архитектуры и моделей разработки ПО Университета Иннополис, m.mazzara@innopolis.ru

**Мавридис Николаос**, руководитель лаборатории когнитивных робототехнических систем Университета Иннополис, n.mavridis@innopolis.ru

Authors:

**Gusmanov Kamill Marselevich**, student of Computer Science, Innopolis University, k.gusmanov@innopolis.ru
**Khanda Kevin Radzhnishevich**, student of Computer Science, Innopolis University, k.khanda@innopolis.ru
**Salikhov Dilshat Bulatovich**, student of Computer Science, Innopolis University, d.salikhov@innopolis.ru
**Mazzara Manuel**, Faculty, Innopolis University, m.mazzara@innopolis.ru
**Mavridis Nikolaos**, Faculty, Innopolis University, n.mavridis@innopolis.ru